\documentstyle[aps,twocolumn]{revtex}
\input BoxedEPS.tex
\SetepsfEPSFSpecial  
\HideDisplacementBoxes
\begin{document}


\def \he {\hbox {$^3$He}}
\def \xe {\hbox {$^{129}$Xe}}
\def \rb {\hbox {$^{85}$Rb}}
\def \h2o {\hbox {H$_2$O}}
\def \td {\hbox {$\tau_{\rm d}$}}
\newcommand{\bohr} {{\rm r}_{{}_{\rm B}}}
\newcommand{\Tt} {\overline{\bf T}}
\newcommand{\Tr} {\hbox {Tr}}
\newcommand{\Eu} {\hbox {Eu}}
\newcommand{\bg} {\begin{equation}}
\newcommand{\ed
} {\end{equation}}
\newcommand{\bzeta} {\bbox{\zeta}}
\newcommand{\Hamip} {\tilde{{\cal H}}_1}     
\newcommand{\roip} {\tilde{\rho}}        
\newcommand{\Ua} {e^{-i{\cal H}_0t}}         
\newcommand{\Ub} {e^{i{\cal H}_0t}}
\newcommand{\drodt} {\frac{d\rho(t)}{dt}}      
\newcommand{\droipdt} {\frac{d\tilde{\rho}(t)}{dt}}  
\newcommand{\Tlong} {\mbox{$T_1$}}
\newcommand{\Ttrans} {\mbox{$T_2$}}
\newcommand {\dQdt} {\frac{d <\tilde{Q}>}{dt}}
\newcommand {\A} {{\cal A}}
\newcommand {\Apq}[2]{A_{#1}^{(#2)}}
\newcommand {\Wpq}[2]{\omega_{#1}^{(#2)}}

\draft

\title{Spin-axis relaxation in spin-exchange collisions of alkali atoms}
\author{S. Kadlecek and T. Walker}
\address{Department of Physics, University of Wisconsin--Madison,
Madison, Wisconsin 53706}
\author{D.K. Walter, C. Erickson and W. Happer}
\address{Department of Physics, Princeton University, Princeton,
New Jersey 08544}
\date{\today}
\maketitle

\begin{abstract} {\it Abstract}: We present calculations of spin-relaxation rates of
alkali-metal atoms due to the spin-axis interaction acting in
binary collisions between the atoms. We show that for the
high-temperature conditions of interest here, the spin relaxation
rates calculated with classical-path trajectories are nearly the
same as those calculated with the distorted-wave Born
approximation. We compare these calculations to recent experiments
that used magnetic decoupling to isolate spin relaxation due to
binary collisions from that due to the formation of triplet
van-der-Waals molecules. The values of the spin-axis coupling
coefficients deduced from measurements of binary collision rates
are consistent with those deduced from molecular decoupling
experiments, and follow a physically plausible scaling law for the spin-axis coupling
coefficients.
\end{abstract}

\narrowtext

\section{Introduction}
\label{sec:intro}

Spin-exchange optical pumping \cite{Carver,RMP} of $^3$He  uses
spin-exchange collisions between $^3$He atoms and optically pumped Rb
atoms
to  produce large
quantities of highly spin-polarized
$^3$He for a variety of applications, including medical imaging
\cite{MacFall} and spin-polarized targets \cite{Anthony}.  The
efficiency
of polarized $^3$He
production is determined by two fundamental rates:  the Rb-He
spin-exchange
rate and the Rb spin-relaxation rate.  The measured spin-exchange rates
 \cite{Baranga}  are in fairly good agreement with
theory\cite{Walker89}.
At the high temperatures needed for the $^3$He
spin-exchange rates to exceed the wall relaxation rates, both Rb-Rb and
Rb-He relaxation limit the efficiency for spin-exchange.  Although it had been  assumed that collisions between
alkali atoms rigorously conserve the spin polarization, Bhaskar et al. \cite{Bhaskar}
discovered that rapid spin-relaxation in fact occurs  in
high-density
optically pumped Cs, with a surprisingly large inferred spin-relaxation
cross section in excess of 1 \AA$^2$. The corresponding cross section
for
Rb-Rb relaxation
\cite{Baranga,Knize,Wagshul,Kadlecek}, while smaller, still limits the
efficiency of $^3$He production.  The yet smaller cross section for K-K
\cite{Knize,KadlecekK} relaxation suggests that, if technical
difficulties
are surmounted, K may be the optimum partner for spin-exchange with
$^3$He\cite{Baranga}.

A few years ago, we discovered that 1/2 to 2/3 of the alkali-alkali
relaxation decouples in magnetic fields of a few kG\cite{Kadlecek},
making implausible the interpretation of the relaxation exclusively
in terms of binary collisions. Recently, careful magnetic decoupling 
studies in low pressure, isotopically pure Rb and Cs samples definitively 
identified the source of the field-dependent relaxation as the spin-axis 
interaction in triplet molecules\cite{Erickson}.  The remaining 
alkali-alkali relaxation mechanism at high magnetic field is then presumably 
from binary collisions.  It is the purpose of this paper to show that the 
deduced values of the spin-axis interaction from binary collisions are
consistent with the values recently obtained from magnetic decoupling 
studies of triplet molecules \cite{Erickson}. We hypothesize a
simple scaling law for the second-order spin-orbit interaction that
explains the relative magnitudes of the observed spin-axis interaction
strengths. Nevertheless, the measured cross sections are in every case 
at least a factor of 10 larger than would be expected from 
{\it ab initio} calculations \cite{Mies}. Table~\ref{table1} contains a 
summary of the existing data on alkali-alkali spin relaxation.

The spin-axis interaction between two alkali-metal atoms is
\bg V_1={2\lambda\over 3}{\bf S}\cdot (3\bzeta
\bzeta-\bbox{1})\cdot {\bf S}~. 
\label{eq:Hamil}
\ed
Here the total electron spin of the valence-electron pair is
$\bf S$ and $\bzeta$ is a unit
vector lying along the direction of the internuclear axis. The coefficient 
$ \lambda(R)$, a rapidly decreasing function of interatomic separation $R$, is currently
believed to arise from both the direct spin-dipolar coupling (averaged over the
electron charge distribution) and the spin-orbit interaction in second
order \cite{Mies}. Accumulating evidence, from both high temperature
\cite{Erickson,Leo} and low temperature experiments\cite{Leo,Chin,Williams}, suggests 
that the predicted spin-axis coupling (presumably arising almost entirely from 
second-order, spin-orbit interactions) is too small in Cs by a factor of 3 to 4, 
and in Rb by a factor of more than 10 \cite{Erickson}. Put another way, the 
theoretical spin-relaxation cross-sections for Rb are smaller than experiment by 
a factor of more than 50.

This paper carefully documents how the collisional averaging of the interaction 
(\ref{eq:Hamil}) leads to a spin relaxation rate. The somewhat complicated 
averaging can be done exactly within the limitations of the classical path
approximation, so the origin of the discrepancy cannot be due to
any inadequacies of the averaging but must lie with the potential
(\ref{eq:Hamil}), or with the spin-independent interatomic potential which
describes the classical paths, or with the neglect of some unknown collisional 
relaxation mechanisms other than sudden binary collisions. For very cold 
collisions, where not so many partial waves are involved, one could use the 
distorted-wave Born approximation (DWBA) \cite{Newbury93} to account for
any limitations of the classical-path method, but at the temperatures of
interest here, as we will show, the results of the classical path 
approximation are within a few percent of those of the DWBA.

Our recent experiments studied the spin-relaxation rates as a
function of magnetic field for the three alkali atoms K, Rb, and Cs, as described
in Refs.~\cite{Kadlecek,Erickson,Kthesis,Ethesis}.  We assume here that the
remaining alkali density dependent contributions to the relaxation rate at
12 kG arise entirely from binary collisions.  The data are summarized in
Table~\ref{table1}.

\section{Classical Paths}
\label{sec:classical}

In a binary collision at temperatures of a few hundred Kelvin many
partial waves contribute to spin-exchange relaxation, so a
classical-path treatment should be adequate. Methods for averaging
over all classical-path cross sections to obtain spin relaxation
rates were given in an earlier paper by Walter {\it et al.}
\cite{Walter} (referred to in the following as WHW) for collisions
between alkali-metal atoms and noble-gas atoms. In particular, the
anistropic magnetic hyperfine interaction between the spin of a
noble gas nucleus and the electron spin of the alkali-metal atom,
Eq. (3) of WHW has the same tensor symmetry as the spin-axis
interaction (\ref{eq:Hamil}), and the detailed calculations are sufficiently similar to those of WHW that we will simply summarize
the results here.

As outlined in WHW and illustrated by Fig. 1 of that paper, we can
assume that the orbit of the colliding pair follows a trajectory
governed by the triplet potential $V_0=V_0(R)$. The
time-dependence of the internuclear separation $R$ can be found
from the equation describing conservation of energy and angular
momentum, 
\bg 
{dR \over dt} = \pm w \sqrt{1-{b^2 \over R^2}-{2 V_0 \over Mw^2}}. 
\label{eq:dRdt} 
\ed 

\noindent
Here $M$ is the reduced mass of the
pair of colliding alkali-metal atoms and $w$ is their relative velocity.  For the 
high-temperature experimental conditions of interest here,
possible changes in direction of the electron spin ${\bf S}$ cause
such small changes in the energy or angular momentum of the
orbital motion that we can neglect them and parametrize the
orbital energy with the initial relative velocity $w$ and the
angular momentum by the impact parameter $b$. 

It is convenient to let the time of closest approach of the pair
be $t=0$, so the orbital angle at time $t$ is \bg
\psi=\psi(t)=\cos^{-1}\bzeta(t)\cdot\bzeta(0). \label {eq:psi} \ed
The time dependence of the orbital angle can be found by numerical
quadrature of the equation for conservation of angular momentum
\bg {d\psi\over dt}= {wb\over R^2}. \label{eq:dpsidt} \ed

When averaged over a thermal distribution of trajectories,
collisions in an alkali-metal vapor of atomic number density $n$
will cause the mean longitudinal electron spin polarization
$\langle S_z\rangle $  of the atoms to relax at the rate 
\bg 
{d \over dt} \langle S_z \rangle =- n \langle v\sigma\rangle \langle
S_z \rangle~~~.
\label{eq:dIbzdt} 
\ed 

\noindent
The rate coefficient can be
readily calculated by methods analogous to those used in WHW for
relaxation due to the anisotropic magnetic dipole hyperfine
interaction.  The average over all angles of the collisions can be
carried out analytically and we find in analogy to (33) of WHW,
\bg \langle v\sigma\rangle =\int_0^{\infty}
dw~p(w)~w~\int_0^{\infty} db~b {8\pi\over
3}\sum_{m=-2}^2|\varphi_{2m}|^2. \label{eq:TpqB} \ed 
The probability $p(w)~dw$ of finding the magnitude $w$ of the relative
velocity of the colliding pair between $w$ and $w+dw$ is 
\bg
p(w)~dw=4\pi w^2\left({M\over 2\pi k_B T}\right )^{3/2} e^{-Mw^2/2k_BT}dw, 
\label{eq:pw} 
\ed 
where $T$ is the absolute temperature
and $k_B$ is Boltzmann's constant. The tensor phases accumulated
during the collision are \bg \varphi_{2m} ={1\over
\hbar}\int_{-\infty}^{\infty} dt~\lambda~d^{2}_{0m}(\psi)
\label{eq:phahfs} \ed Here $d^2_{0m}(\psi)$ is a Wigner $d$
function, for example, $d^2_{00}(\psi)=(3\cos^2\psi-1)/2$ 
\cite{Varshalovich}. In the
integrand of Eq.~(\ref{eq:phahfs}), both the spin-axis coupling
coefficient $\lambda$ and the orbital angle $\psi$ are functions
of time $t$, obtained from numerical integration of Equations
(\ref{eq:dRdt}) and (\ref{eq:dpsidt}). Since $\psi(-t)=-\psi(t)$,
$d_{2m}(\psi)$ is an even function of $t$ if $m$ is even and odd
if $m$ is odd. Also $\lambda$ is an even function of $t$
(measured from the time of closest approach), so $\phi_{2m}$ is
identically zero if $m=\pm 1$. In practice, the rapid decrease of
$\lambda$ with increasing internuclear separation $r$ means that
$\phi_{2,\pm 2}$ is so small as to be negligible in practice and so the $m=0$ contribution to (\ref{eq:TpqB}) dominates.

\section{Partial Waves}
\label{sec:partial-waves}

We can also calculate the spin relaxation due to the spin-axis
interaction (\ref{eq:Hamil}) by the distorted-wave Born
approximation (DWBA), outlined by Newbury {\it et al.} \cite{Newbury93},
which we refer to as Newbury in this section. According to (47) of
Newbury, the rate coefficient corresponding to (\ref{eq:TpqB}) is
\bg \langle v\sigma\rangle = \int_0^{\infty} dw~p(w)~w~ {64 \pi
M^2\over 3\hbar^4k^6}\sum_{l=0}^\infty(2l+1)\lambda_{ll}^2
\label{eq:vsigdwba} \ed For the high-temperature conditions of
interest here the three matrix elements $\lambda_{ll'}$ for
spin-flip scattering from an initial wave of angular momentum $l$
to a final wave of angular momentum $l'=l,l\pm 2$ are nearly
equal. We have assumed exact equality to reduce the double sum on
$l$ and $l'$ to a single sum on $l$ in (\ref{eq:vsigdwba}).

The classical initial relative velocity $w$ is related to the
asymptotic spatial frequency $k$ of the scattered wave by \bg
w={\hbar k\over M} \label{eq:wtok} \ed In (\ref{eq:vsigdwba}), 
the integral over impact
parameters $b$ which occurs in the classical-path expression
(\ref{eq:TpqB}) is replaced by a sum over partial waves $l$. The
matrix element of the spin-axis coupling coefficient between
partial waves $l$ and $l'$ is 
\bg 
\lambda_{ll'} = k\int_0^{\infty}g_l(r)\lambda(r)g_{l'}(r)~dr.
\label{eq:lamllp} 
\ed 

\noindent The wave functions $g_l(r)$ are solutions of the 
Schr\"odinger equation 
\bg 
\left (-{d^2\over dr^2}+{l(l+1)\over r^2}+{2M\over
\hbar^2}V_0-k^2\right)g_l=0, 
\label{eq:scheq} 
\ed 

\noindent
with the boundary condition as $r \rightarrow 0$ \bg g_l\to 0 \ed and with the
asymptotic boundary condition for $kr/l\to \infty$ 
\bg 
g_l \to \sin\left(kr-{\pi l\over 2}+\delta_l\right).
\ed
From comparison of (\ref{eq:TpqB}) with (\ref{eq:vsigdwba}) we
conclude that the classical-path and partial wave treatment will give
practically the same answers if 
\bg
\sum_{m=-2}^2|\varphi_{2m}|^2=\left( 2 \lambda_{ll}\over{E}\right)^2. \label{eq:dwbapw} 
\label{eq:comparison}
\ed 

\noindent
In (\ref{eq:dwbapw}) we have
assumed that $l=kb$. The initial relative energy of the colliding
pair is 
\bg 
E={\hbar^2k^2\over 2M} \label{eq:E} 
\ed
Numerical solutions to the differential equations (\ref{eq:dRdt})
and (\ref{eq:scheq}) are readily obtained. We have confirmed that
relation (\ref{eq:comparison}) is indeed true, establishing the 
equivalence of the classical-path and partial-wave methods of calculating
the spin-relaxation rate coefficients. Although the two methods give
the same results, the classical-path approach is much less numerically
intensive, requiring only the solution of the simple first-order differential
equations (\ref{eq:dRdt}) and (\ref{eq:dpsidt}), whereas the partial-wave
analysis requires solving the second-order equation (\ref{eq:scheq}), with
rapidly oscillating solutions.

\section{Comparison of ab initio calculations to experiment}\label{expt}

In the context of ultracold collisions, Mies {\it et al.}
\cite{Mies} recently published {\it ab initio} calculations of
$\lambda$, as the sum of two contributions \bg \lambda =
\lambda_{{}_{\rm SO}} + \lambda_{{}_{\rm DD}}.
\label{eq:lamSO-lamDD} \ed For the heavier alkali-metal atoms,
and for small $R$, they
found that second-order spin-orbit interactions, represented in
(\ref{eq:lamSO-lamDD}) by $\lambda_{{}_{\rm SO}}$, were much larger
than the term $\lambda_{{}_{\rm DD}}$, which describes the
direct interaction between the magnetic moments of the two
valence electrons. Second-order contributions analogous to those
responsible for $\lambda_{{}_{\rm SO}}$ are well known from the
theoretical literature on molecular spectroscopy \cite{Tinkham,Julienne}. 
Mies {\it et al.} parameterize their calculations of $\lambda$ as follows 
(we have converted their results from atomic units):
\bg \lambda ={3g_S^2
\mu_B^2\over 4 a_B^3} \left[ C e^{-\beta(R-R_S)}-\left ({a_B\over
R}\right )^3\right].
\label{eq:jul}
\ed
Here $a_B$ is the Bohr radius, $\mu_B$ is the Bohr magneton and
$g_S=2.00232$ is the electronic g-factor. The results (\ref{eq:jul}) of
the
{\it ab initio} calculations are parametrized as follows: for
Rb, $R_S=5.292 $ \AA , $C=.001252$ and $\beta=1.84$ \AA $^{-1}$;
for Cs, $R_S=5.292$ \AA , $C=.02249$ and $\beta=1.568$ \AA
$^{-1}$. The first term in (\ref{eq:jul}) represents
$\lambda_{{}_{\rm SO}}$. The second term, 
\bg
\lambda_{{}_{\rm DD}} = - {3 g_S^2 \mu_B^2 \over 4 R^3},
\label{eq:lamDD}
\ed
represents the magnetic interaction of electrons, taken as point particles 
separated by a distance $R$. This is an excellent approximation at very 
large $R$, but should be modified, as we show below, at 
smaller values of $R$ where the most important contributions to spin 
relaxation occur.  This modification, however, is unlikely to have a major impact on the predictions of Ref.~\cite{Mies}.

Table~\ref{table1} shows cross-sections calculated as described
in Sections \ref{sec:classical} and \ref{sec:partial-waves}. For K, we assume
only the classical spin-dipolar term, because the spin-orbit
contribution
estimated by
scaling from Mies {\it et al.} \cite{Mies} is negligible.  As can be
seen
from Table~\ref{table1}, the theoretical estimates are smaller than experiment by
about a factor
of 10 for Cs and K, and a factor of almost 60 for Rb  where the {\it ab
initio}
calculations
predict that
$\lambda$ goes to zero at $R=5.5$ \AA.

In order to describe the effects of spin-relaxation in a number of
experiments on Cs, where (\ref{eq:jul}) predicts relaxation rates that
are much too small, the NIST group has chosen to multiply
$\lambda_{{}_{\rm SO}}$
(the computed contribution to $\lambda $ from second-order spin-orbit
interactions) by a constant value. This assumes that the
$R$-dependence of the calculation is correct\cite{Leo,Williams}.
Guided by new experimental data, we will discuss similar scaling
arguments in Section~\ref{sec:scaling}.


\section{Effects of wavefunction overlap on $\lambda_{{}_{\rm DD}}$}
\label{overlap}

The expression (\ref{eq:lamDD}) of $\lambda_{{}_{\rm DD}}$
 neglects the spatial distribution of the electron
charge. In this section we present a simple estimate of the effect
of the spatial distribution, and find that its neglect 
cannot be responsible for the discrepancy between experiment and theory.

A simple estimate of the wavefunction 
of the triplet state of an alkali dimer is
\begin{eqnarray}
|\Psi \rangle &=&\psi({\bf r}_1,{\bf r}_2)|\chi \rangle \nonumber\\
&=&N[\varphi_A(1)\varphi_B(2)-\varphi_B(1)\varphi_A(2)]\;|\chi \rangle
\label{eq:lcao}
\end{eqnarray}

\noindent
where, for example, $\varphi_A(1)$ is a spatial orbital for electron 1 centered at
nucleus A, $N$ a normalizing factor, and $|\chi \rangle$ is a three-component
spinor representing the triplet spin state. The matrix element
between a final triplet state $|\Psi_f\rangle$
and an initial triplet state $|\Psi_i\rangle$ of the electronic magnetic
dipole interaction (at fixed $R$), is
\begin{eqnarray}
\langle \Psi_f| {g_S^2\mu_B^2 \over r_{12}^5}{\bf S}_1 \cdot (r_{12}^2\bbox{1}-3{\bf r}_{12}{\bf r}_{12})
\cdot {\bf S}_2 |\Psi_i\rangle \nonumber \\
= {4 \lambda_{{}_{\rm DD}} \over 3} \langle\chi_f| {\bf S}_1\cdot (3 \bzeta\bzeta-\bbox{1})\cdot{\bf S}_2|\chi_i\rangle
\end{eqnarray}
where
\begin{equation}\lambda_{{}_{\rm DD}}= {3g_S^2\mu_B^2\over4}\int d^3r_1
d^3r_2
{r_{12}^2-3z_{12}^2\over 2 r_{12}^5}|\psi ({\bf r}_1,{\bf r}_2)|^2
\end{equation}
We make the simplifying assumption that $\varphi_A$ and
$\varphi_B$ can be
approximated by the ground-state wavefunction of the valence electron
of an isolated alkali-metal atom.

\begin{figure}
\BoxedEPSF{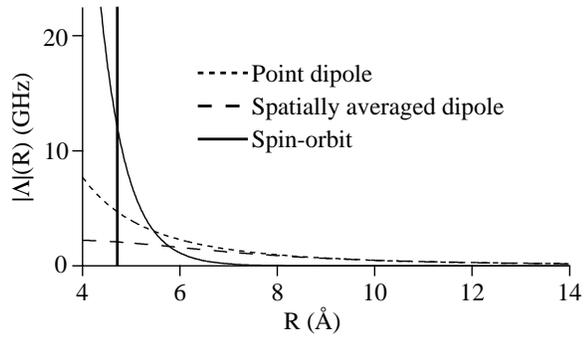 scaled 520}
\caption{{\it Ab initio}
calculations of $\lambda_{{}_{\rm DD}}$ for Rb from Ref.~\protect\cite{Mies},
and the modified
version obtained by
spatially averaging the magnetic dipole-dipole contribution as described
in
the text.  The
vertical solid line is at the classical turning point for a zero-impact
parameter collision at
500 K collision energy.}\label{figdip}\end{figure}

The results are shown in Fig.~\ref{figdip}.  The principle effect of the
wavefunction overlap is to reduce the value of $\lambda_{{}_{\rm DD}}$
as compared to the
point dipole
approximation, and to reduce the predicted {\it ab initio} cross section
for
Rb-Rb to $5.3\times 10^{-20}$ cm$^2$, increasing the experiment/theory
discrepency.
Use of  better electron wave functions
than the simple form (\ref{eq:lcao}) are unlikely to change the results 
by the orders of magnitude needed to obtain agreement with experiment.

\section{Scaling Relation for $\lambda_{{}_{\rm SO}}$}
\label{sec:scaling}

In this section we show that the spin-axis coupling coefficients deduced
from high temperature experiments on triplet molecules are consistent with a simple and plausible
scaling
relation that, in turn, accurately predicts the relative binary
spin-relaxation cross sections.
Clearly, the contribution $ \lambda_{{}_{\rm DD}}$ from the direct
interaction of the magnetic dipole moments of the electrons is
much too small to account for observed relaxation rates in the
heavier alkali-metal atoms. The {\it ab initio} calculation of the additional contribution
$\lambda_{{}_{\rm SO}}$ is also too
small. In order to test the consistency of the observed molecular and binary relaxation rates,  we shall
assume that Mies {\it et al.} have correctly identified
the second-order spin-orbit interaction as a major contributor
to $\lambda$, which implies that $\lambda_{{}_{\rm SO}}$ should be 
proportional to the square of the $P_{1/2}$--$P_{3/2}$ fine structure
splitting $\Delta \nu$, and inversely proportional to the valence-electron
binding energy $E$, as predicted by perturbation theory. To obtain the
radial dependence of $\lambda_{{}_{\rm SO}}$, we shall make the physically
plausible assumption that it scales as $|\phi(r)|^2$, the valence electron 
density of an unperturbed alkali-metal atom at a distance $r$ from the nucleus.
(The radial dependence of the values of $\lambda_{{}_{\rm SO}}$ calculated by 
Mies {\it et al.} is very nearly that of $ |\phi(r)|^2$.) Thus our scaling law is

\begin{equation}
\lambda_{{}_{\rm SO}} = \Omega \frac{(h\Delta \nu)^2}{E} |\phi(r)|^2.
\label{eq:lamscale}
\end{equation}
The fine structure splittings $\Delta \nu/c$ for Cs, Rb, and K are 
$554~{\rm cm}^{-1}$, $237.6~{\rm cm}^{-1}$, and $57.7~{\rm cm}^{-1}$, respectively;
the binding energies $E$ are
$3.89~{\rm eV}$, $4.18~{\rm eV}$, and $4.34~{\rm eV}$. 

For the required wavefunctions $\phi(r)$, we use the asymptotic expansion of the 
Coulomb wave function \cite{Bates49}, namely
\bg
\phi(r) = {N \over r} \left(\frac{2r}{n^*}\right)^{n^*} e^{-r/n^*}~~~,
\ed
where the radius $r$ is measured in Bohr radii $a_B$. The effective principal quantum 
number $n^*$ of the valence electron
in its ground state is related to the ionization energy $E$ (in eV) and the Rydberg 
$R_{\infty} = 13.61~{\rm eV}$ by $n^* = (R_{\infty}/E)^{1/2}$, and the normalization
factor $N$ is given by $N=[(4\pi)^{1/2}(n^*)^{3/2}\Gamma(n^*)]^{-1}$, where 
$\Gamma(x)$ is the Euler gamma function.

The universal constant $\Omega$ of (\ref{eq:lamscale}), which has units of volume, 
is deduced from experiments as follows. Recent experimental studies \cite{Erickson} 
of the Cs spin relaxation in triplet dimers yield a spin-axis coupling 
$|\lambda^{\rm Cs}/h| = 2.79~{\rm GHz}$. This value reflects a thermal average over the 
rovibrational states of the triplet molecules, but for simplicity we will take it to be 
the value of $\lambda/h$ at the Cs triplet dimer equilibrium internuclear separation of 
$R=12~a_B$, whence, by Eqs.~(\ref{eq:lamSO-lamDD}) and (\ref{eq:lamDD}),
$\lambda_{{}_{\rm SO}}^{\rm Cs} = 2.94 ~{\rm GHz}$. Thus, using
$|\phi|^2 = 3.48 \times 10^{-6}~a_B^{-3}$ at $R=12~a_B$, we obtain
\bg
{\Omega} = 2880~a_B^3~~~,
\ed
roughly an atomic volume.

In Rb, the equilibrium internuclear separation for triplet molecules is $R=11.5~a_B$, 
where $|\phi|^2= 3.23 \times 10^{-6}~a_B^{-3}$, whence Eqs.~(\ref{eq:lamscale}) and
(\ref{eq:lamDD}) yield 
$\lambda^{\rm Rb}/h = 294~\hbox{MHz}$, in close agreement with the value
$|\lambda^{\rm Rb}/h| = 290~\hbox{MHz}$ deduced from the observation of
spin-relaxation due to Rb triplet dimers \cite{Erickson}.

Using the results of Sections \ref{sec:classical} and \ref{sec:partial-waves} and the 
spin-axis coupling coefficients deduced from the above simple arguments, we may readily 
compute the spin relaxation due to the spin-axis interaction in binary collisions between
alkali-metal atoms. The calculated cross sections for K-K, Rb-Rb and Cs-Cs at 400 K, using
the well-known {\it ab initio} interatomic potentials of Krauss and
Stevens\cite{Krauss} to compute the needed trajectories, are
respectively
$4.4\times 10^{-20}~{\rm cm}^2$, $1.4 \times 10^{-18}~{\rm cm}^2$ and
$1.1
\times 10^{-16}~{\rm cm}^2$.  As can be seen from Table~\ref{table1}, these values
agree with experiment to within a factor of 3 for rubidium and cesium. We note that for potassium
it is the dipole-dipole interaction that dominates the
contribution
to $\lambda$, and so the larger discrepancy with experiment for K is
may be
due to the simplified estimate (\ref{eq:lamDD}) of $\lambda_{{}_{\rm DD}}$.

\section{Conclusions}

We have shown here that the strength of the spin-axis coupling deduced
from measurements of binary collisions between alkali atoms is consistent
with the values deduced from magnetic decoupling of relaxation due to formation 
of weakly bound triplet molecules.  The deduced spin-axis coupling strengths 
are much larger than predicted by {\it ab initio} theory.  Clearly more reliable
theoretical estimates of the spin-axis coupling are needed.

\section{Acknowledgements}
Support for this work came from the National Science Foundation, AFOSR,
and DARPA.  D.K.W. is supported by the Hertz Foundation.  We appreciate
helpful discussions with P. Leo, P. Julienne, and C. Williams.

\begin{table}[tb]
\caption{Spin-relaxation cross sections $\langle v\sigma \rangle/ \bar v$ (in $10^{-18}$ cm$^2$) for
alkali atoms. Zero-field ``cross sections'' include contributions from both molecular formation
and binary collisions. High-field cross sections are assumed to arise solely from binary
collisions. The theoretical values are calculated, at a temperature of 400 K, using
estimates of the spin-axis coupling strength either from a simple scaling law or from 
Ref.~\protect\cite{Mies}. Also shown is the ratio of experiment  
to {\it ab initio} predictions.}
\label{table1}
\begin{tabular}{lcccccr}
    &           	& Zero-field        	& High-field        	& Scaling      		& {\it ab initio}   	&    \\
Atom& Ref.  		& Expt.        	& Expt.        	& Theory        	& Theory            	&Ratio\\ \hline
K   & \cite{Knize}    	& $2.4$  &           		&          		&           		&    \\
K   & \cite{Kthesis}  	& $1.0$  & $0.62$  & $0.044$ &$0.067$ 	& 9.2\\\hline 
Rb  & \cite{Knize}    	& $16$  &           	 &          		&           		&    \\ 
Rb  & \cite{Wagshul}  	&$18$  &           	 &          		&           		&    \\ 
Rb  & \cite{Baranga}  	& $9.2$&           	 &          		&     			&    \\ 
Rb  & \cite{Kadlecek} 	& $9.3$ & $3.4$  &$1.4 $ & $0.061$  & 56 \\ 
Rb  & \cite{Ethesis}  	& $15$  &$5.6$  & $1.4 $ & $0.061$  & 92 \\ \hline 
Cs  &\cite{Bhaskar}  	& $203$ &           		&          		&           		&    \\ 
Cs  &\cite{Leo}	& $230$  & $110$  & $110$ & $11$ & 10 \\
\end{tabular}
\end{table}

\end{document}